# Near-surface effects in modelling oscillations of $\eta$ Boo

Jørgen Christensen-Dalsgaard

*Teoretisk Astrofysik Center, Danmarks Grundforskningsfond*

Timothy R. Bedding

*School of Physics, University of Sydney 2006, Australia*

Günter Houdek

*Institut für Astronomie, Universität Wien, Austria*

Hans Kjeldsen, Colin Rosenthal, Regner Trampedach

*Teoretisk Astrofysik Center, Danmarks Grundforskningsfond*

Mario J.P.F.G. Monteiro

*Centro de Astrofisica, Universidade do Porto, Portugal*

Åke Nordlund

*Niels Bohr Institutet, Københavns Universitet, Denmark*

Following the report of solar-like oscillations in the G0 V star $\eta$ Boo (Kjeldsen et al. 1995), a first attempt to model the observed frequencies was made by Christensen-Dalsgaard et al. (1995). This attempt succeeded in reproducing the observed frequency separations $\Delta\nu$ and $\delta\nu_{02}$, although there remained a difference of $\sim 10\,\mu$Hz between observed and computed frequencies. In those models, the near-surface region of the star was treated rather crudely: convection was described by means of a local mixing-length theory neglecting turbulent pressure, and the oscillations were assumed to be adiabatic. These approximations are likely to affect both $\Delta\nu$ and absolute frequencies.

Here we consider more sophisticated models of the external regions:

- **NL.a** A non-local mixing-length theory (Balmforth 1992) including turbulent pressure. Frequencies computed in the adiabatic approximation.
- **NL.na** The same, but with nonadiabatic frequencies, including the Lagrangian turbulent-pressure perturbation.
- **HYD.a** Outer structure obtained from horizontal and temporal averaging of a hydrodynamical convection model (e.g. Stein & Nordlund 1991). This was extended into a deep-envelope model through continuous matching at the lower boundary of the hydrodynamical model. Frequencies computed in the adiabatic approximation.

We compared the radial oscillation frequencies computed from these sophisticated models with those of normal models whose (local) mixing length was adjusted to obtain the same convection-zone depth. The comparisons are shown in the figure, in the sense (sophisticated treatment) − (normal treatment). Note that oscillations in $\eta$ Boo were observed in the frequency range 750–950 $\mu$Hz.

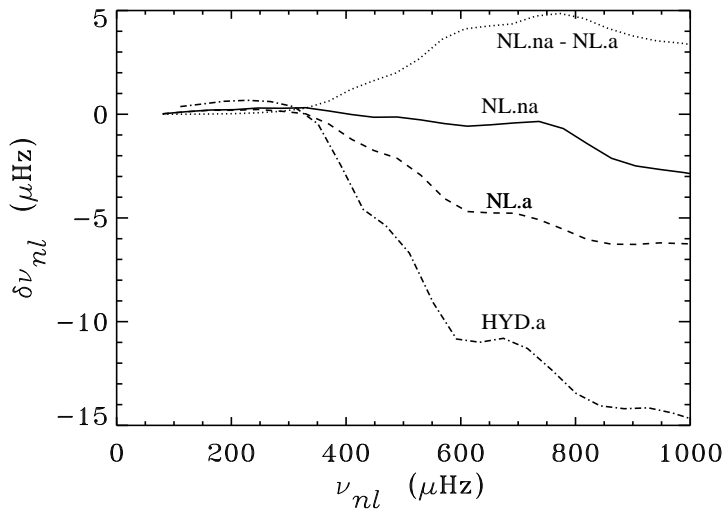

Two effects influence the frequencies: (i) Changes in the equilibrium model structure. It can be shown that these are dominated by the inclusion of turbulent pressure. Our results for **NL.a** and **HYD.a** imply frequency shifts of −5 to −15 $\mu$Hz in the region of the observed oscillations, with the region of substantial turbulent pressure being considerably larger in the **HYD** model. (ii) The physics of the oscillations (nonadiabaticity and turbulent-pressure perturbations), whose effects are shown by the dotted line (**NL.na** − **NL.a**). Interestingly, the two influences almost cancel for the **NL** model (solid line), implying that little correction is required to the frequencies of Christensen-Dalsgaard et al. (1995). However, the **HYD.a** result suggests that the **NL** calculations underestimate the change in the equilibrium structure.

The corrections to the large frequency separation depend on the slope of the curve where the oscillations are observed. From the figure we estimate corrections to $\Delta\nu$ of up to about $-1\,\mu$Hz.

It is clear that assumptions in modelling the near-surface region introduce significant uncertainties in the computed frequencies. The present results give some hope that more detailed observations of $\eta$ Boo, the Sun and other stars will help improve our understanding of the physics of convection.